\documentstyle[preprint,prb,aps]{revtex}
\begin{document}
\def\be{\begin{equation}}
\def\ee{\end{equation}}
\def\ba{\begin{eqnarray}}
\def\ea{\end{eqnarray}}
\draft
\title{ Thermal and Mechanical Properties of some FCC Transition  
Metals and their Binary  Alloys} 
\author{T. \c{C}a\u{g}{\i}n}
\address{Materials and Process Simulation Center, California  Institute of Technology, Pasadena,}
\address{CA91125 , U.S.A.}
\author{G. Dereli, M. Uludo\u{g}an, M. Tomak}
\address{Department of Physics, Middle East Technical University, 
06531 Ankara, TURKEY}  
\date{\today}
\maketitle
\begin{abstract}
\noindent 
The temperature dependence of thermodynamic and mechanical properties of 
six fcc transition metals (Ni, Cu, Ag, Au, Pt, Rh) and the alloying 
behavior of Ag-Au and Cu-Ni are studied using  molecular dynamics (MD). 
The structures are described at elevated temperatures by the force fields 
developed by Sutton and 
co-workers within the context of tight binding approach. MD algorithms 
are based on the  extended Hamiltonian 				
formalism from the works of Andersen, Parinello and Rahman, Nos\'{e},
Hoover and \c{C}a\u{g}{\i}n. The SIMULATOR program  
that we use generates information about various
physical properties during the run time along with
critical trajectory and stepwise information
which need to be analyzed post production. The thermodynamic and
mechanical properties are calculated in the temperature range between  
300K to 1500K with 200K increments using the statistical
fluctuation 
expressions over the MD trajectories.

\end{abstract}
\section{Introduction}
The theoretical and computational modeling is becoming increasingly important 
in the development of advanced high performance materials for industrial
applications.
Atomistic level understanding of the properties
of fcc transition metals
under various conditions is important in their technological
applications. 
The behavior of pure silver, gold, copper, nickel, platinum, rhodium and their
alloys are tested over a wide temperature range using the potentials developed by
Sutton and co-workers \cite{1,2} for simulation. Our computer simulation
results for Pt-Rh alloys are presented in elsewhere\cite{3}.
The MD algorithms that we use are based on the extended Hamiltonian
formalism and the ordinary experimental conditions are simulated
using the constant-pressure, constant-temperature (NPT) MD method.

Computer simulations on various model systems usually use simple
pair potentials. On many occasions to account for the directionality
of bonding, three body interactions are also employed. 
The interactions in real crystalline materials can not be represented by
simple pairwise interactions alone. Pure pairwise potential model gives the
Cauchy relation between the elastic constants $C_{12}=C_{44}$,which is not   
the case in real metals. In these systems the  
electron density plays a dominant role in interactions and resulting 
physical properties. Therefore, many-body interactions should be taken 
into consideration in any study of  metals and metal alloys. 
In simple sp-bonded metals, the interaction potentials may be derived 
from model pseudopotentials using the second order perturbation theory. 
 We have developed interaction potentials along these lines,
and utilized them to study the properties of simple alkali metal and alkali 
metal alloys in the past \cite{4}.  However, for d-band metal and metal
 alloys the model pseudopotential approach should give way to newer 
 techniques evolved over the past ten years to account for the many body
  effects.  Among these approaches, the empirical many body potentials based on 
Norskov's Effective Medium Theory \cite{5}, Daw and Baskes' Embedded Atom 
Method \cite{6}, Finnis and Sinclair's \cite{7,8} empirical many body
potentials, and more recently the many body potentials developed by 
Sutton and co-workers \cite{1,2} within the context of tight binding 
approach \cite{9} for fcc transition metals can be listed.
Due to its mathematically simple power law form and fairly long range
character, in recent years the Sutton-Chen (SC) potential has been widely
used in simulations to study a range of problems \cite{10,11,12,13,14,15,16}. 
 
\section{ The Method}
In the SC description the total potential energy of the metal  
is given as a sum of a pairwise repulsion
term and a many-body density dependent cohesion term. The cohesion term
supplies the  description of short-range interactions to obtain
a good description of surface relaxation, and the pairwise term gives a
correct description of long-range interactions with a van der Waals   
tail. The functional form of the interaction potential is as follows:
\be
U_i = D\left ({1\over 2}  \sum_j u({\bf r_{ij}})  - c \rho_i \right)
\label{must24} ~~,
\ee
where
\ba
u (r) = ({a \over r} )^n  \label{must25} ~~, \\
\rho_i = \left(\sum_j \phi(r_{ij}) \right)^{1\over 2} \label{must26}~~, \\
\phi(r) = \left({a \over r} \right)^m  \label{must27}~~.
\ea

The Sutton-Chen potential parameters $D$, $c$ and $m$ and $n$ are
optimized to fit to the 0 K properties such as the cohesive energy,
zero pressure condition and the bulk modulus of the f.c.c. metals. In
Table:1, the values of these parameters for Ni, Cu, Ag, Au, Pt 
and Rh are listed.

The functional form of the Sutton-Chen potential is fairly simple in 
comparison to Embedded Atom Method potentials and is moderately long 
ranged.  The last property makes this set especially attractive for 
surface and interface studies amongst
others, since most of them are very short ranged (are fitted up to
first or second nearest neighbor distances).These interaction 
potentials can be generalized to describe binary   
metal alloys in such a way that all the parameters in the Hamiltonian
equations are obtained from the parameters of pure metals. The
Sutton-Chen interaction potential above is adopted by Rafii-Tabar and  
Sutton \cite{2} to a random f.c.c. alloy model in which sites are
occupied by  two types of atoms completely randomly, such that the alloy
has the required average concentration. The equilibrium lattice
parameter $a^*$ at 0 K of the random alloy is chosen as the
universal length scale and the expectation value $E^t$ per atom of the
interaction Hamiltonian is given as a function of $a^*$. Rafii-Tabar and
Sutton after determining the value of the equilibrium lattice parameter
for the random alloy,  they calculated elastic constants and enthalpy of
mixing by {\it static} lattice summation method. Once $a^*$ is found the
enthalpy of mixing $\Delta H$ per atom at 0 K are also obtained from
\be
\Delta H= E^t - c_A E^A  - c_B E^B \label{must28} ~~,
\ee
where $E^A$ and $E^B$ are the cohesive energies per atom of the
elemental $A$ and $B$ metals and the constants are such that $c_A + c_B
=1$.

In this paper we aim at using the Molecular Dynamics method to obtain the 
$a^*$ and the enthalpy of mixing per atom for alloys. In the following the
expressions specific to many body potentials  
which are used in our computations are presented.The many body force on 
atom $a$ along a direction $i(=x,y,z)$ is given as:
\be
F_{ia}= -{D \over 2}\left (\sum_b^*  u'(r) {r_{abi}\over r_{ab}}  -   
{c_a\over
2} {\sum_b^* \phi'(r){r_{abi} \over r_{ab} } \over \rho_a}\right ),
\label{musta1}
\ee
where $'$ denotes $ {\partial \over \partial r}$ and $^*$ signifies the
exclusion of $a=b$ from the sums. The anisotropic stress tensor
including the contribution from the many body potential is calculated
from
\be
\Omega P_{ij}= \sum_a{ p_{ai} p_{aj} \over m_a }- {D\over 2}  \sum_a   
\left (
\sum_b^*  u'(r) {r_{abi} r_{abj}\over r_{ab}}  - {c_a\over 2} {\sum_b^*
\phi'(r){ r_{abi} r_{abj}\over r_{ab} } \over \rho_a}\right),
\label{musta2}
\ee
The potential energy contribution to the elastic constants, the
hypervirial tensor $\chi_{ijkl}$, is given as
\ba
\Omega \chi_{ijkl} &=& {D\over 2}  \sum_a ( \sum_b^* ( u'' - {u'   
\over r_{ab} } ){r_{abi} r_{abj}r_{abk} r_{abl}\over r^2_{ab}} \nonumber
\\ 
& & - {c_a \over 2}  {\sum_b^* (\phi'' - {\phi' \over r_{ab} }) {r_{abi}
r_{abj}r_{abk} r_{abl}\over r^2_{ab}} \over \rho_i} \label{musta3} \\
& & + {c_a \over 4} { (\sum_b^* \phi' { r_{abi} r_{abj}\over r_{ab} } )
(\sum_c^* \phi' { r_{ack} r_{acl} \over r_{ac} } ) \over \rho_a^3
}) . \nonumber
\ea
In our computations at each concentration and at each temperature, first
the
zero strain state ,$h_o$, of the system is determined  by performing
constant
temperature and constant stress simulations (NPT)at zero stress. This
yields the reference shape and size matrix, $h_o$ in
Parrinello-Rahman formalism. In determining elastic constants this
reference state is used in constant temperature constant volume
simulations (NVE) of 50000 steps for each state point. The elastic
constants
are evaluated using the following statistical fluctuation formulas \cite{17}
\ba
C^T_{ijkl} &=& - {\Omega_o \over k_B T } (<P_{ij} P_{kl}> -
<P_{ij}><P_{kl}>) \nonumber \\
& & + {2Nk_B T (\delta_{ik} \delta_{jl} + \delta_{il} \delta_{jk} )\over
\Omega_o} +  <\chi_{ijkl}> \label{musta4}
\ea
where $ < > $ denotes the averaging over time and $\Omega_o = det h_o $
is the reference volume for the model system. The first term represents
the contribution from the fluctuation of the microscopic stress tensor, 
P$_{ij}$, the second term represents the kinetic energy contribution,
and the third term is the Born term.

\vskip  4mm

We use the program {\bf Simulator} developed by \c Ca\u g\i n  
that employs the state of the art MD algorithms based on
extended Hamiltonian formalisms emerging from the works of
Anderson \cite{18}, Parinello and Rahman \cite{19}, Nos\'{e} \cite{20}, 
Hoover \cite{21} and \c{C}a\u{g}{\i}n \cite{22}.
A 500 atoms cubical system is used and the simulation started 
with atoms randomly
distributed on a fcc lattice.The system is thermalized starting from 1K
to the target temperature using a constant enthalpy and constant pressure
(NHP) ensemble by slowly heating while scaling velocities to increment
the temperature of 1K/step over the specific number of steps depending on
the target temperature. 
This is followed by strict velocity scaling at each target temperature.
Then, NPT dynamics at this temperature for 
20000 steps to calculate the volume, density and enthalpy of the system
for each concentration is performed.  The resulting zero strain averaged matrix 
$<h_0>$  is used in calculating elastic constants over 50000 steps 
of NVE dynamics.
A fifth-order Gear predictor-corrector algorithm is used in $\Delta$t = 2 
fs.The Parrinello-Rahman piston mass parameter is chosen as W=400 and in 
NPT runs the Nos\'{e}-Hoover parameter is set to Q=100.    

\section{ Results and Discussion}

In this section,the molecular simulation results obtained for pure f.c.c.
transition metals, Ni, Cu, Ag, Au, Pt and Rh are presented.
In Tables 2 to 7,
the results of NPT molecular 
dynamics
simulations of pure elements after 20000 steps are shown. These are
density, enthalpy, potential energy and volume of 500 atoms 
calculated  
at different temperatures. 
In obtaining these values the system is thermalized starting from
$1$K to the target temperature using  20000 steps NHP simulations.
We then performed another 20000 steps NPT dynamics at the target
temperature. 
In order to show the  behavior of 
Sutton-Chen potentials at elevated temperatures; the results of density and  
enthalpy are plotted at temperatures  $300$K, $500$K, $750$K and
$1000$K in Figures 1 and 2. The extra datas  for Pt and
Rh at
temperatures $700$K, $900$K, $1100$K, $1300$K and $1500$K
are coming from the calculations of Pt-Rh alloys \cite{3}.
In Tables 2 to 7
at $300$K, the experimental and simulation results of the densities
are shown. Simulation results 
show approximately 2.0$\%$
deviation from the
experimental values for Ni,Cu,Ag,Au and 1.4$\%$
 and 1.0$\%$ for Pt and Rh
respectively.
 The percent change in the lattice parameter at
each temperature (the lattice parameter at $300$K is used as the
reference) for Ni, Cu , Ag, Au
are given in Table 8,  
 Pt and Rh results are being discussed in \cite{3}.
 As a concluding remark, it can be said that as the temperature increases
the deviation from the experimental values increases.

In Tables 9 to 14  
, the results of the calculations of the
elastic constants and bulk moduli of Ni, Cu, Ag, Au, Pt and Rh obtained
from NVE simulations of 50000 steps are presented.  
Elastic constants are calculated in the temperature range between $300$K
to $1000$K. 
In the calculations at each concentration and at each temperature  
first the zero-strain state of the system is determined by performing 
constant-temperature and constant-stress simulations. In determining the
elastic constants this reference state is used in constant-temperature, constant-volume simulations of 50000 steps for each state point. 
Comparison with experimental results is possible only at $100$K and this is shown in Table 15. 
The ratio of $C_{12}/C_{44}$ is $2.38$ for Ni,
$2.42$ for Cu, $1.75$ for Ag, $3.76$ for Au, $3.68$ for Ph and $1.69$ for Rh.
Figures 3 to 8 show the variations of elastic constants
of Ni, Cu, Ag, Au, Pt, Rh with respect to temperature.
The change in the bulk moduli of these metals by heating is shown
in Figure 9 .
The simulation values of the bulk moduli at $1000$K are the 
fingerprints for the
melting temperature of these metals. The bulk modulus is higher
for higher melting temperature.
The elastic constants results show that the crystals are 
elastically
stable since the stability conditions $C_{44}>0$, $C_{11}>0$ and
$C_{11}>C_{12}$ are satisfied and  thermal softening behavior is observed 
as the temperature is increased.

In Tables 16 and 17, the simulation results of 
Ag-Au and Cu-Ni binary alloys are given. These are the
enthalpy of mixing and the densities of 
Ag-Au and Cu-Ni alloys at $300$K.
In Figure 10, the densities of Cu-Ni and Ag-Au are drawn with respect to 
atomic concentrations. Again in Figure 11, the enthalpy of mixing for the
same alloys are drawn as a function of atomic concentrations. The sign of
enthalpy of mixing is correct at all concentration values. 
The mixing of these alloys are enthalpically favorable.
The potentials used  in the present dynamic simulations  give 
reasonably accurate description of the thermodynamic properties
and the elastic constants.
Altough, the parametrization of these potentials are based on the  bulk
properties at  $0$K, still it can describe the temperature-dependent 
behavior of the solid correctly. We find that with an improved parameter
sets the Sutton-Chen potential does quite well in predicting a number of
properties of fcc metals in MD simulations. In order to simulate the
ordinary experimental conditions (constant pressure) the NPT MD method
works quite well and determining the zero strain state, $h_o$,
of the system at each  concentration and at each temperature  
is curicial in simulations.

\vskip 6mm
\acknowledgements{

This research is supported by the Scientific and Technical Research Council 
of Turkey (TUBITAK) through the Project TBAG 1592 and by the Middle East
Technical University Research Fund through the Project AFP-96-01-05-03.
}

\begin{figure}
\caption{Densities of Ni,Cu,Ag,Au,Pt and Rh as a function of temperature}
\end{figure}
\begin{figure}
\caption{Enthalpies of Ni,Cu,Ag,Au,Pt and Rh as a function 
of temperature}
\end{figure}
\begin{figure}
\caption{Elastic constants of Ni as a function of temperature}
\end{figure}
\begin{figure}
\caption{Elastic constants of Cu as a function of temperature}
\end{figure}
\begin{figure}
\caption{Elastic constants of Ag as a function of temperature}
\end{figure}
\begin{figure}
\caption{Elastic constants of Au as a function of temperature}
\end{figure}
\begin{figure}
\caption{Elastic constants of Pt as a function of temperature}
\end{figure}
\begin{figure}
\caption{Elastic constants of Rh as a function of temperature}
\end{figure}
\begin{figure}
\caption{Bulk moduli of Ni,Cu,Ag,Au,Pt and Rh as a function of temperature}
\end{figure}
\begin{figure}
\caption{Densities of Ag-Au and Cu-Ni as a function of concentration at $300$K}
\end{figure}
\begin{figure}
\caption{Enthalpies of mixing of Ag-Au and Cu-Ni as a function of concentration at $300$K}
\end{figure}

\begin{table}
\caption{\label{tab1}
 Sutton-Chen potential parameters for f.c.c. metals.}
\begin{center}
\begin{tabular}{|lcccccc|}
\hline 
 & & & & & & \\
a  & $D$ & c  & m & n &  Metal &\\
 $(\AA)$ & $(10^{-2} eV)$& & & & & \\
\hline
&&&&&&\\
3.52 & 1.57070 & ~39.432 & 6 & ~9 & Ni & \\
3.61 & 1.23820 & ~39.432 & 6 & ~9 & Cu & \\
4.09 & 0.25415 & 144.410 & 6 & 12 & Ag & \\
4.08 & 1.27930 & ~34.408 & 8 & 10 & Au & \\
3.92 & 1.98330 & ~34.408 & 8 & 10 & Pt & \\
3.80 & 0.49371 & 144.410 & 6 & 12 & Rh & \\
\hline
\end{tabular}  
\end{center}
\end{table}

\begin{table}
\caption{\label{tab2}
The density, enthalpy, potential energy and volume of Ni as obtained   from NPT molecular dynamics simulations after 20000 steps. The number in
parenthesis is the corresponding experimental value at that
temperature.}
\begin{center}
\begin{tabular}{|lccccc|}
\hline
&&&&&\\
~~~~T  &~~~~~~~~ $\rho$~~~~~~~~ &~~~~~ H~~~~  &~~~~~ U~~~~ &~500 x $V_0$~ &
\\
~~ (K) & $(g/cm^3)$&$(kJ/mol)$&$(kJ/mol)$&$(nm^3)$& \\
\hline
&&&&&\\
~300 & 8.7468 (8.90)  & -420.77492 & -424.51844 & 5.571 &  \\
~500 & 8.6327 ~~~~~~~ & -415.54424 & -421.78116 & 5.645 &  \\
~750 & 8.4815 ~~~~~~~ & -408.77260 & -418.12898 & 5.745 &  \\
1000 & 8.3176 ~~~~~~~ & -401.66660 & -414.13714 & 5.858 &  \\
\hline
\end{tabular}
\end{center}
\end{table}

\begin{table}
\caption{\label{tab3}
The density, enthalpy, potential energy and volume of Cu as obtained
from NPT molecular dynamics simulations after 20000 steps. The number in 
parenthesis is the corresponding experimental value at that
temperature. }
\begin{center}
\begin{tabular}{|lccccc|}
\hline
&&&&&\\
~~~~T  &~~~~~~~~ $\rho$~~~~~~~~ &~~~~~ H~~~~  &~~~~~ U~~~~ &~500 x $V_0$~ &
\\
~~ (K) & $(g/cm^3)$&$(kJ/mol)$&$(kJ/mol)$&$(nm^3)$& \\
\hline 
&&&&&\\
~300 &  8.7235 (8.912)   & -330.03778 & -333.77963 & 6.048 &  \\
~500 &  8.5742 ~~~~~~~~~~& -324.71396 & -330.95169 & 6.153 &  \\
~750 &  8.3717 ~~~~~~~~~~& -317.73297 & -327.08560 & 6.302 &  \\
1000 &  8.1334 ~~~~~~~~~~& -310.01874 & -322.48972 & 6.486 &  \\
\hline 
\end{tabular}
\end{center}
\end{table}

\begin{table} 
\caption{\label{tab4}
The density, enthalpy, potential energy and volume 
of Ag as obtained from NPT molecular dynamics simulations after 20000 
steps. The number in parenthesis is the corresponding experimental value 
at that temperature.} 
\begin{center}
\begin{tabular}{|lccccc|}
\hline 
&&&&&\\
~~~~T  &~~~~~~~~ $\rho$~~~~~~~~ &~~~~~ H~~~~  &~~~~~ U~~~~ &~500 x $V_0$~ &
\\
~~ (K) & $(g/cm^3)$&$(kJ/mol)$&$(kJ/mol)$&$(nm^3)$& \\
\hline
&&&&&\\
~300 & 10.2710 (10.49)   & -277.95578 & -281.69931 & 8.719 &  \\
~500 & 10.1030 ~~~~~~~~~~& -272.64944 & -278.88458 & 8.864 &  \\
~750 & ~9.8755 ~~~~~~~~~~& -265.69861 & -275.05511 & 9.068 &  \\
1000 & ~9.6191 ~~~~~~~~~~& -258.25937 & -270.73138 & 9.310 &  \\
\hline
\end{tabular}
\end{center}
\end{table}

\begin{table} 
\caption{\label{tab5}
The density, enthalpy, potential energy and volume of Au as obtained
from NPT molecular dynamics simulations after 20000 steps. The number in
parenthesis is the corresponding experimental value at that
temperature.}
\begin{center}
\begin{tabular}{|lccccc|}
\hline 
&&&&&\\
~~~~T  &~~~~~~~~ $\rho$~~~~~~~~ &~~~~~ H~~~~  &~~~~~ U~~~~ &~500 x $V_0$~ &
\\
~~ (K) & $(g/cm^3)$&$(kJ/mol)$&$(kJ/mol)$&$(nm^3)$& \\
\hline
&&&&&\\
~300 & 18.8411 (19.32)   & -357.05188 & -360.79401 & 8.679 &  \\
~500 & 18.5168 ~~~~~~~~~~& -351.69836 & -357.93332 & 8.831 &  \\
~750 & 18.0595 ~~~~~~~~~~& -344.58054 & -353.93341 & 9.055 &  \\
1000 & 17.4804 ~~~~~~~~~~& -336.48160 & -348.95288 & 9.355 &  \\
\hline
\end{tabular}
\end{center}
\end{table}

\begin{table}
\caption{\label{tab6}
The density, enthalpy, potential energy and volume of Pt as obtained
from NPT molecular dynamics simulations after 20000 steps. The number in
parenthesis is the corresponding experimental value at that
temperature.}
\begin{center}
\begin{tabular}{|lccccc|}
\hline
&&&&&\\
~~~T  &~~~~~~~~ $\rho$~~~~~~~~ &~~~~~ H~~~~  &~~~~~ U~~~~ &~500 x $V_0$~ &
\\
~~ (K) & $(g/cm^3)$&$(kJ/mol)$&$(kJ/mol)$&$(nm^3)$& \\
\hline
&&&&&\\
~300 & 21.1954 (21.50)   & -557.82990 & -561.57330 & 7.641 &  \\
~500 & 20.9757 ~~~~~~~~~~& -552.62994 & -558.86822 & 7.722 &  \\
~700 & 20.7427 ~~~~~~~~~~& -547.30029 & -556.03040 & 7.808 &  \\
~750 & 20.6825 ~~~~~~~~~~& -545.93384 & -555.28748 & 7.831 &  \\
~900 & 20.4961 ~~~~~~~~~~& -541.79657 & -553.02002 & 7.902 &  \\
1000 & 20.3664 ~~~~~~~~~~& -538.96735 & -551.43976 & 7.952 &  \\
1100 & 20.2303 ~~~~~~~~~~& -536.06519 & -549.78400 & 8.006 &  \\
1300 & 19.9378 ~~~~~~~~~~& -530.03790 & -546.25262 & 8.123 &  \\
1500 & 19.5992 ~~~~~~~~~~& -523.49762 & -542.20453 & 8.264 &  \\
\hline
\end{tabular}
\end{center}
\end{table}

\begin{table}
\caption{\label{tab7}
The density, enthalpy, potential energy and volume of Rh as obtained
from NPT molecular dynamics simulations after 20000 steps. The number in
parenthesis is the corresponding experimental value at that
temperature.}
\begin{center}
\begin{tabular}{|lccccc|}
\hline
&&&&&\\
~~~T  &~~~~~~~~ $\rho$~~~~~~~~ &~~~~~ H~~~~  &~~~~~ U~~~~ &~500 x $V_0$~ &
\\
~~ (K) & $(g/cm^3)$&$(kJ/mol)$&$(kJ/mol)$&$(nm^3)$& \\
\hline
&&&&&\\
~300 & 12.3152 (12.45)   & -547.24719 & -550.99084 & 6.937 & \\
~500 & 12.2184 ~~~~~~~~~~& -542.09912 & -548.33862 & 6.992 & \\
~700 & 12.1191 ~~~~~~~~~~& -536.88055 & -545.61493 & 7.049 & \\
~750 & 12.0937 ~~~~~~~~~~& -535.56250 & -544.91638 & 7.064 & \\
~900 & 12.0162 ~~~~~~~~~~& -531.56805 & -542.78949 & 7.110 & \\
1000 & 11.9637 ~~~~~~~~~~& -528.87659 & -541.34656 & 7.141 & \\
1100 & 11.9099 ~~~~~~~~~~& -526.14417 & -539.86707 & 7.173 & \\
1300 & 11.7990 ~~~~~~~~~~& -520.59625 & -536.80499 & 7.241 & \\
1500 & 11.6838 ~~~~~~~~~~& -514.93768 & -533.62976 & 7.312 & \\
\hline
\end{tabular}
\end{center}
\vspace{2cm}
\end{table}
\vspace{2cm}

\begin{table}
\caption{\label{tab8}
The linear expansion of Ni, Cu, Ag, and Au as a
function of
temperature.}
\begin{center}
\begin{tabular}{|c|c|c|c|c|}
\hline
 \multicolumn{2} {|c|}{T~(K)}  & 500 & 750 & 1000 \\
\hline
Ni & This Work & 0.44 & 1.04 & 1.72 \\ \cline{2-2}
& Experiment & 0.29 & 0.69 & 1.13  \\
\hline
Cu & This Work & 0.58 & 1.40 & 2.42 \\ \cline{2-2}
& Experiment & 0.34 & 0.82 & 1.36  \\
\hline
Ag & This Work & 0.55 & 1.34 & 2.26 \\ \cline{2-2}
& Experiment & 0.43  & 0.99 & 1.61 \\
\hline
Au & This Work & 0.58  & 1.44  & 2.60 \\ \cline{2-2}
& Experiment & 0.31 & 0.70 & 1.14 \\
\hline  
\end{tabular}
\end{center}
\end{table}

\begin{table}
\caption{
 Elastic constants and bulk modulus of Ni calculated at 300 K, 500
K, 750 K and 1000 K as obtained from NVE Molecular Dynamics
Simulation after 50000 steps.
\label{tab9}}
\begin{center}
\begin{tabular}{|lccccc|}
\hline
&&&&&\\
~~~~T  &~~~~~~ $C_{11}$~~~~~~ &~~~~ $C_{12}$~~~~  &~~~~ $C_{44}$~~~~
&~~~~~$B$~~~~~~& \\
~ (K) & (GPa) & (GPa) & (GPa) & (GPa) & \\
\hline
~300 & 213.76~~~$\pm$0.19 & 166.57~~~$\pm$0.20 & ~69.77~~~$\pm$0.21 &
182.30~~~$\pm$0.20 & \\
~500 & 201.70~~~$\pm$0.21 & 159.48~~~$\pm$0.51 & ~63.56~~~$\pm$0.20 &
173.55~~~$\pm$0.41 & \\
~750 & 186.05~~~$\pm$0.66 & 150.34~~~$\pm$0.06 & ~55.94~~~$\pm$0.55 &   
162.24~~~$\pm$0.26 & \\
1000 & 169.14~~~$\pm$1.99 & 141.08~~~$\pm$0.87 & ~48.29~~~$\pm$0.28 & 
150.43~~~$\pm$1.24 & \\
\hline
\end{tabular}
\end{center}
\end{table}

\begin{table}
\caption{\label{tab10}
Elastic constants and bulk modulus of Cu calculated at 300 K, 500
K, 750 K and 1000 K as obtained from NVE Molecular Dynamics
Simulation after 50000 steps.}
\begin{center}
\begin{tabular}{|lccccc|}
\hline
&&&&&\\
~~~~T  &~~~~~~ $C_{11}$~~~~~~ &~~~~ $C_{12}$~~~~  &~~~~ $C_{44}$~~~~
&~~~~~$B$~~~~~~& \\
~ (K) & (GPa) & (GPa) & (GPa) & (GPa) & \\
\hline
~300 & 153.06~~~$\pm$0.42 & 119.45~~~$\pm$0.16 & ~49.35~~~$\pm$0.27 &
130.65~~~$\pm$0.25 & \\
~500 & 140.87~~~$\pm$0.19 & 112.78~~~$\pm$0.24 & ~43.46~~~$\pm$0.40 &
122.14~~~$\pm$0.23 & \\
~750 & 126.46~~~$\pm$1.06 & 104.43~~~$\pm$0.32 & ~36.37~~~$\pm$1.39 &
111.77~~~$\pm$0.57 & \\
1000 & 109.09~~~$\pm$0.66 & ~94.40~~~$\pm$0.94 & ~27.92~~~$\pm$0.65 &
~99.30~~~$\pm$0.85 & \\
\hline
\end{tabular}
\end{center}
\end{table}

\begin{table}
\caption{\label{tab11}
Elastic constants and bulk modulus of Ag calculated at 300 K, 500
K, 750 K and 1000 K as obtained from NVE Molecular Dynamics
Simulation after 50000 steps.}
\begin{center}
\begin{tabular}{|lccccc|}
\hline
&&&&&\\
~~~~T  &~~~~~~ $C_{11}$~~~~~~ &~~~~ $C_{12}$~~~~  &~~~~ $C_{44}$~~~~
&~~~~~$B$~~~~~~& \\
~ (K) & (GPa) & (GPa) & (GPa) & (GPa) & \\
\hline
~300 & 126.95~~~$\pm$0.23 & ~88.49~~~$\pm$0.11 & ~50.50~~~$\pm$0.08 &
101.31~~~$\pm$0.15 & \\
~500 & 117.46~~~$\pm$0.27 & ~83.45~~~$\pm$0.13 & ~45.50~~~$\pm$0.28 &
~94.79~~~$\pm$0.17 & \\
~750 & 104.16~~~$\pm$0.58 & ~76.79~~~$\pm$0.72 & ~37.75~~~$\pm$0.33 &
~85.91~~~$\pm$0.68 & \\
1000 & ~91.88~~~$\pm$0.86 & ~70.59~~~$\pm$1.27 & ~31.54~~~$\pm$0.80 &
~77.69~~~$\pm$1.14 & \\
\hline
\end{tabular}
\end{center}
\end{table}

\begin{table}
\caption{\label{tab12}
Elastic constants and bulk modulus of Au calculated at 300 K, 500
K, 750 K and 1000 K as obtained from NVE Molecular Dynamics
Simulation after 50000 steps.}
\begin{center}
\begin{tabular}{|lccccc|}
\hline
&&&&&\\
~~~~T  &~~~~~~ $C_{11}$~~~~~~ &~~~~ $C_{12}$~~~~  &~~~~ $C_{44}$~~~~
&~~~~~$B$~~~~~~& \\
~ (K) & (GPa) & (GPa) & (GPa) & (GPa) & \\
\hline
~300 & 158.24~~~$\pm$0.64 & 131.56~~~$\pm$0.02 & ~34.92~~~$\pm$0.10 &
140.45~~~$\pm$0.23 & \\
~500 & 141.74~~~$\pm$0.75 & 118.92~~~$\pm$0.21 & ~30.08~~~$\pm$0.56 &
126.52~~~$\pm$0.39 & \\
~750 & 123.92~~~$\pm$0.62 & 107.42~~~$\pm$0.92 & ~24.10~~~$\pm$0.58 &
112.92~~~$\pm$0.82 & \\
1000 & ~95.43~~~$\pm$0.86 & ~84.78~~~$\pm$0.16 & ~16.78~~~$\pm$0.81 &
~88.33~~~$\pm$0.39 & \\
\hline
\end{tabular}
\end{center}
\end{table}

\begin{table}
\caption{\label{tab13} 
Elastic constants and bulk modulus of Pt calculated from 300 K to
1500
K with 200 K increments as obtained from NVE Molecular Dynamics
Simulation after 50000 steps.}
\begin{center}
\begin{tabular}{|lccccc|}
\hline
&&&&&\\
~~~~T  &~~~~~~ $C_{11}$~~~~~~ &~~~~ $C_{12}$~~~~  &~~~~ $C_{44}$~~~~
&~~~~~$B$~~~~~~& \\
~ (K) & (GPa) & (GPa) & (GPa) & (GPa) & \\
\hline
~300 & 289.63~~~$\pm$1.02 & 239.55~~~$\pm$0.16 & ~65.07~~~$\pm$0.28 &
256.24~~~$\pm$0.45 & \\
~500 & 272.67~~~$\pm$0.87 & 227.11~~~$\pm$0.77 & ~59.67~~~$\pm$0.36 &
242.29~~~$\pm$0.80 & \\
~700 & 256.12~~~$\pm$1.71 & 214.38~~~$\pm$0.78 & ~54.15~~~$\pm$0.59 &
228.30~~~$\pm$1.09 & \\
~750 & 251.09~~~$\pm$1.06 & 211.05~~~$\pm$0.72 & ~52.90~~~$\pm$0.28 &
224.34~~~$\pm$0.84 & \\
~900 & 233.28~~~$\pm$0.40 & 196.58~~~$\pm$0.32 & ~49.08~~~$\pm$0.50 &
208.81~~~$\pm$0.35 & \\
1000 & 227.47~~~$\pm$2.66 & 193.46~~~$\pm$0.81 & ~45.50~~~$\pm$0.50 &
204.80~~~$\pm$1.43 & \\
1100 & 219.39~~~$\pm$0.80 & 188.60~~~$\pm$0.62 & ~44.12~~~$\pm$1.51 &
198.86~~~$\pm$0.68 & \\
1300 & 200.63~~~$\pm$2.39 & 174.16~~~$\pm$0.51 & ~38.78~~~$\pm$0.55 &
182.98~~~$\pm$1.14 & \\
1500 & 177.28~~~$\pm$2.46 & 155.50~~~$\pm$0.40 & ~31.94~~~$\pm$1.96 &
162.76~~~$\pm$1.09 & \\
\hline
\end{tabular}
\end{center}
\end{table}

\begin{table}
\caption{\label{tab14}
Elastic constants and bulk modulus of Rh calculated from 300 K to
1500
K with 200 K increments as obtained from NVE Molecular Dynamics
Simulation after 50000 steps.}
\begin{center}
\begin{tabular}{|lccccc|}
\hline
&&&&&\\
~~~~T  &~~~~~~ $C_{11}$~~~~~~ &~~~~ $C_{12}$~~~~  &~~~~ $C_{44}$~~~~
&~~~~~$B$~~~~~~& \\
~ (K) & (GPa) & (GPa) & (GPa) & (GPa) & \\
\hline
~300 & 322.30~~~$\pm$0.66 & 223.02~~~$\pm$0.90 & 131.95~~~$\pm$0.69 &
256.11~~~$\pm$0.82 & \\
~500 & 312.03~~~$\pm$0.69 & 216.89~~~$\pm$0.44 & 124.72~~~$\pm$0.15 &
248.60~~~$\pm$0.52 & \\
~700 & 298.92~~~$\pm$0.77 & 210.23~~~$\pm$0.78 & 117.44~~~$\pm$0.14 &
239.79~~~$\pm$0.78 & \\
~750 & 295.14~~~$\pm$0.88 & 207.72~~~$\pm$0.73 & 116.27~~~$\pm$1.27 &
236.86~~~$\pm$0.78 & \\
~900 & 286.44~~~$\pm$0.62 & 203.55~~~$\pm$0.69 & 110.80~~~$\pm$0.97 &
231.17~~~$\pm$0.67 & \\
1000 & 279.36~~~$\pm$1.12 & 199.07~~~$\pm$0.71 & 108.16~~~$\pm$1.07 &
225.83~~~$\pm$0.84 & \\
1100 & 274.83~~~$\pm$1.49 & 197.32~~~$\pm$0.20 & 104.07~~~$\pm$0.10 &
223.15~~~$\pm$0.63 & \\
1300 & 263.48~~~$\pm$0.77 & 191.06~~~$\pm$2.16 & ~97.36~~~$\pm$0.88 &
215.20~~~$\pm$1.70 & \\
1500 & 249.59~~~$\pm$0.12 & 184.75~~~$\pm$1.18 & ~91.29~~~$\pm$1.32 &
206.36~~~$\pm$0.83 & \\
\hline
\end{tabular}
\end{center}
\end{table}

\begin{table}
\caption{\label{tab12}
Elastic constants  of fcc metals in units of GPa at $300$K.
At each entry, the first number gives the MD simulation result while the second
number in round brackets is the experimental value.}

\begin{center}
\begin{tabular}{|lccc|}
\hline
&&&\\
~~~~Metal  &~~~~~~ $C_{11}$~~~~~~ &~~~~ $C_{12}$~~~~  &~~~~ $C_{44}$~~~~ \\
\hline
~Ni & 213.76 (250.8) & 166.50 (150.0) & ~69.77 (123.5)  \\

~Cu & 153.06 (168.39) & 119.45 (121.42) & ~49.35 (~75.39)  \\

~Ag & 126.95 (123.99) & ~88.49 (~93.67) & ~50.50 (~46.12)  \\

~Au & 158.24 (192.34) & 131.56 (163.14) & ~34.92 (~41.95)  \\
\hline
\end{tabular}
\end{center}
\end{table}

\begin{table}
\caption{\label{tab15} 
The enthalpy of mixing and density for the random 
Ag-Au binary alloy at T= 300 K as obtained from TPN molecular dynamics 
simulations after 20000 to 25000 steps.}
\begin{center}
\begin{tabular}{|cccc|}
\hline
&&&\\
Percent & ~~~~~~~~ $\rho$~~~~~~~~ &~~~~~ H~~~~  &~~~~~~$\Delta$ H~~~~~ \\
Au @ Ag &  $(g/cm^3)$&$(kJ/mol)$&$(J/mol)$ \\ 
\hline
&&&\\
~~0 &10.2708 & -277.95450 & 0.0 \\
~10 &11.1411 & -286.73332 & -869.67 \\
~20 &12.0089 & -295.38254 & -1609.74 \\
~30 &12.8757 & -303.84702 & -2165.07 \\
~40 &13.7408 & -312.16146 & -2570.36 \\
~50 &14.6009 & -320.20978 & -2709.53 \\
~60 &15.4590 & -328.08450 & -2675.10 \\
~70 &16.3107 & -335.64754 & -2328.99 \\
~80 &17.1593 & -343.42174 & -2193.44 \\
~90 &18.0029 & -350.15168 & -1014.83 \\
100 &18.8409 & -357.04600 & 0.0\\
\hline
\end{tabular}
\end{center}
\end{table}

\begin{table}
\caption{\label{tab16} 
 The enthalpy of mixing and density for the random
Cu-Ni binary alloy at T= 300 K as obtained from TPN molecular dynamics
simulations after 20000 to 25000 steps.}
\begin{center}
\begin{tabular}{|cccc|}
\hline
&&&\\
Percent & ~~~~~~~~ $\rho$~~~~~~~~ &~~~~~ H~~~~  &~~~~~~$\Delta$ H~~~~~ \\
Ni @ Cu &  $(g/cm^3)$&$(kJ/mol)$&$(J/mol)$ \\
\hline
&&&\\
~~0 & 8.7241 & -330.03506 & 0.0 \\
~10 & 8.7368 & -339.24688 & -137.13 \\
~20 & 8.7490 & -348.45416 & -269.72 \\
~30 & 8.7560 & -357.59454 & -335.41 \\
~40 & 8.7629 & -366.73834 & -410.14 \\
~50 & 8.7667 & -375.82918 & -410.67 \\
~60 & 8.7663 & -384.89876 & -416.70 \\
~70 & 8.7650 & -393.93110 & -374.56 \\
~80 & 8.7611 & -402.92374 & -291.16 \\
~90 & 8.7550 & -411.86898 & -161.71 \\
100 & 8.7691 & -420.78196 & 0.0 \\
\hline
\end{tabular}
\end{center}
\end{table}

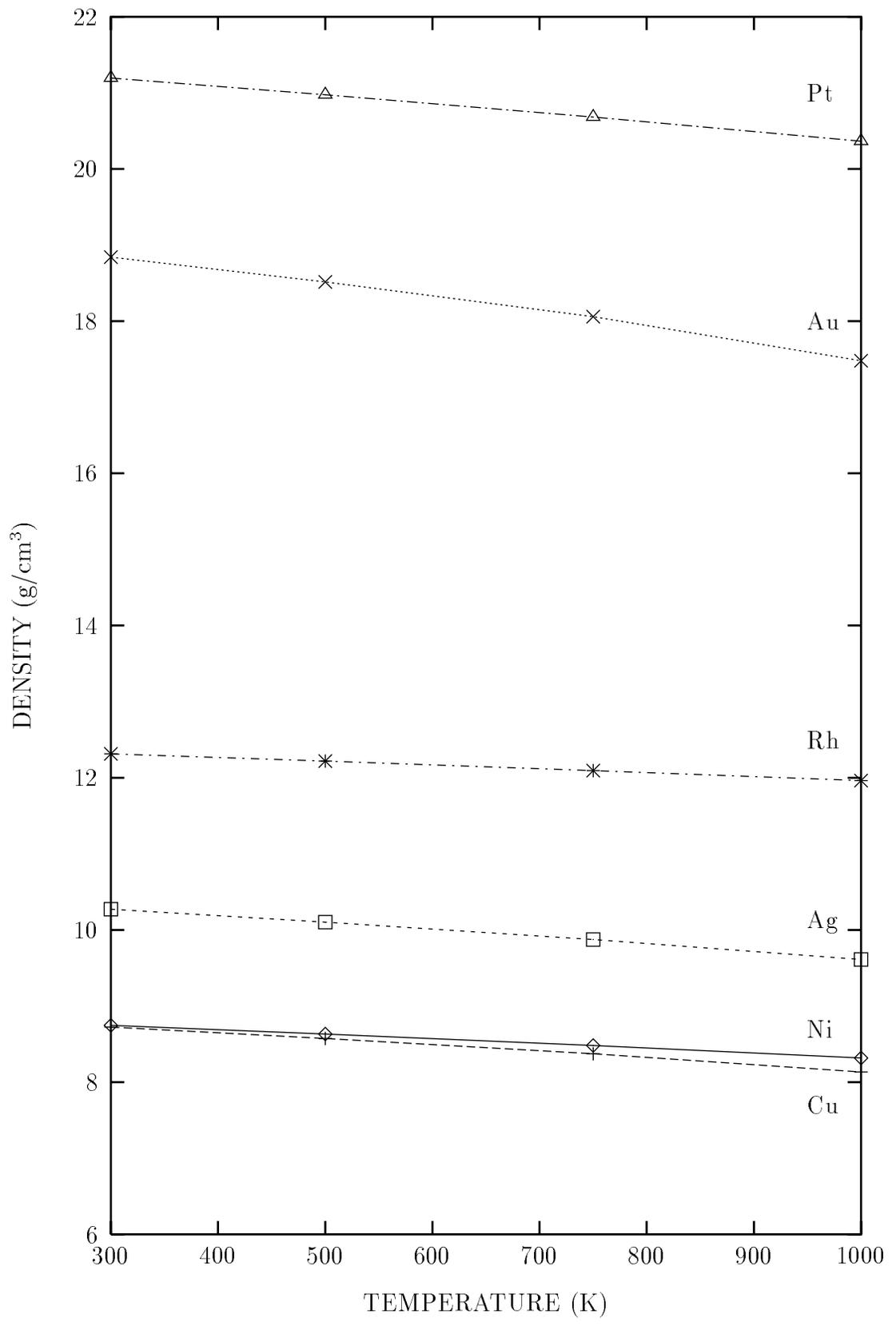
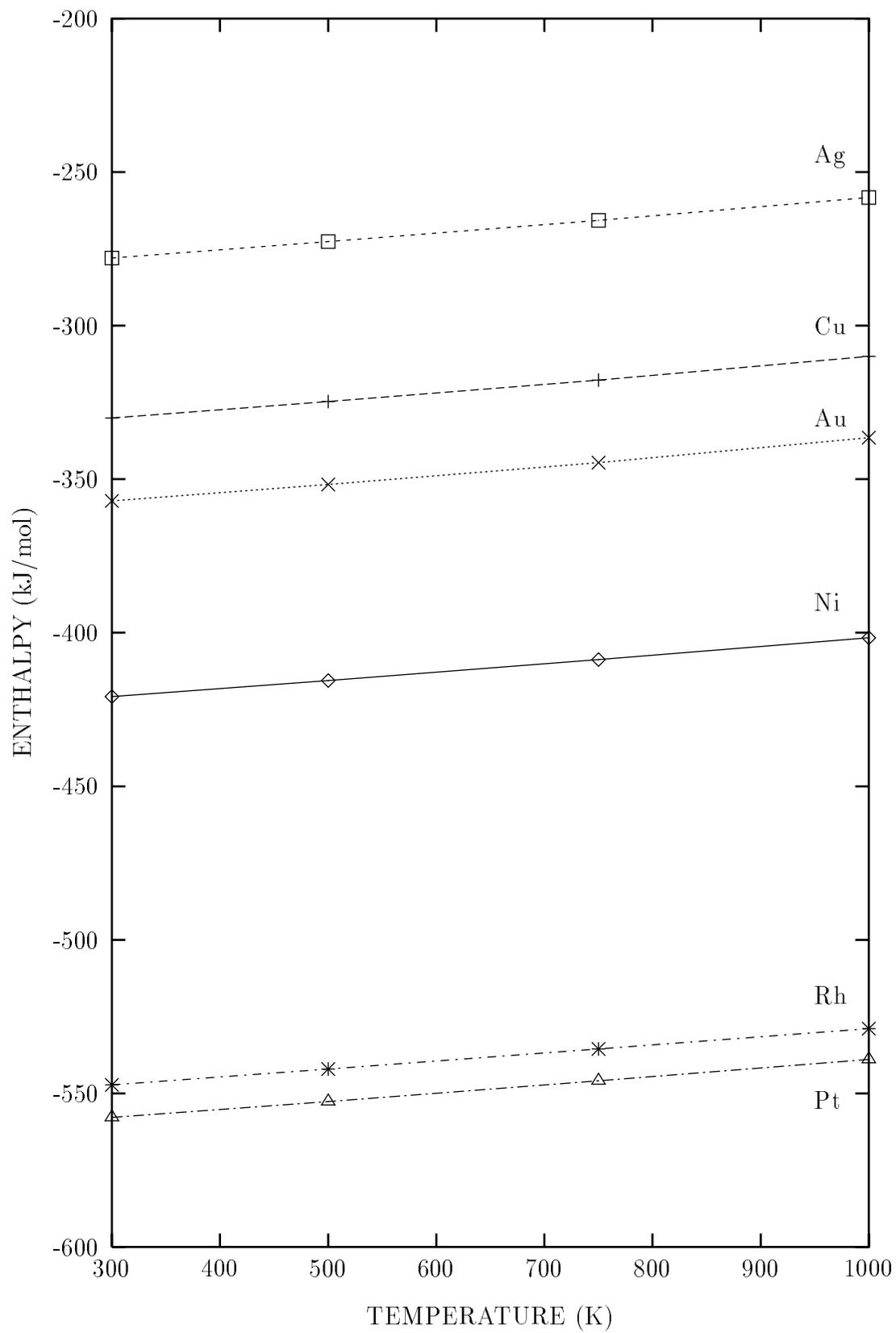
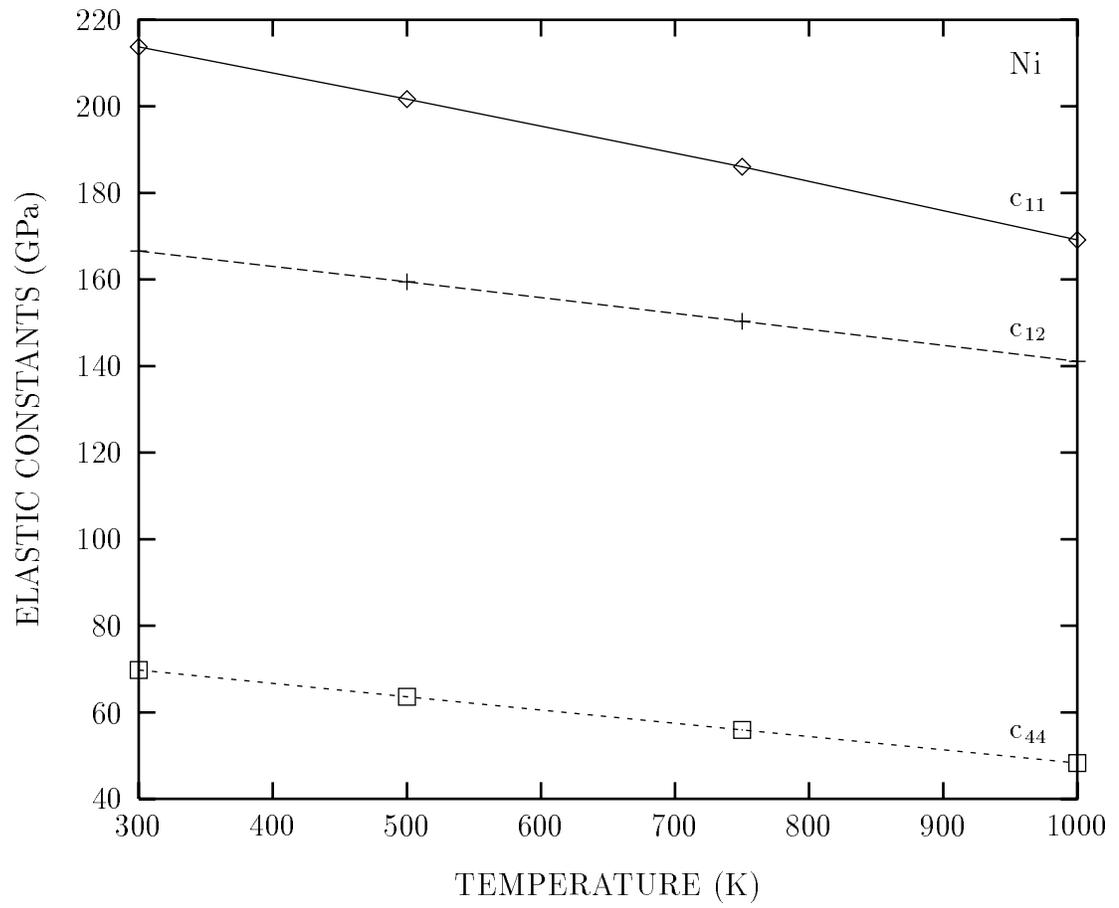
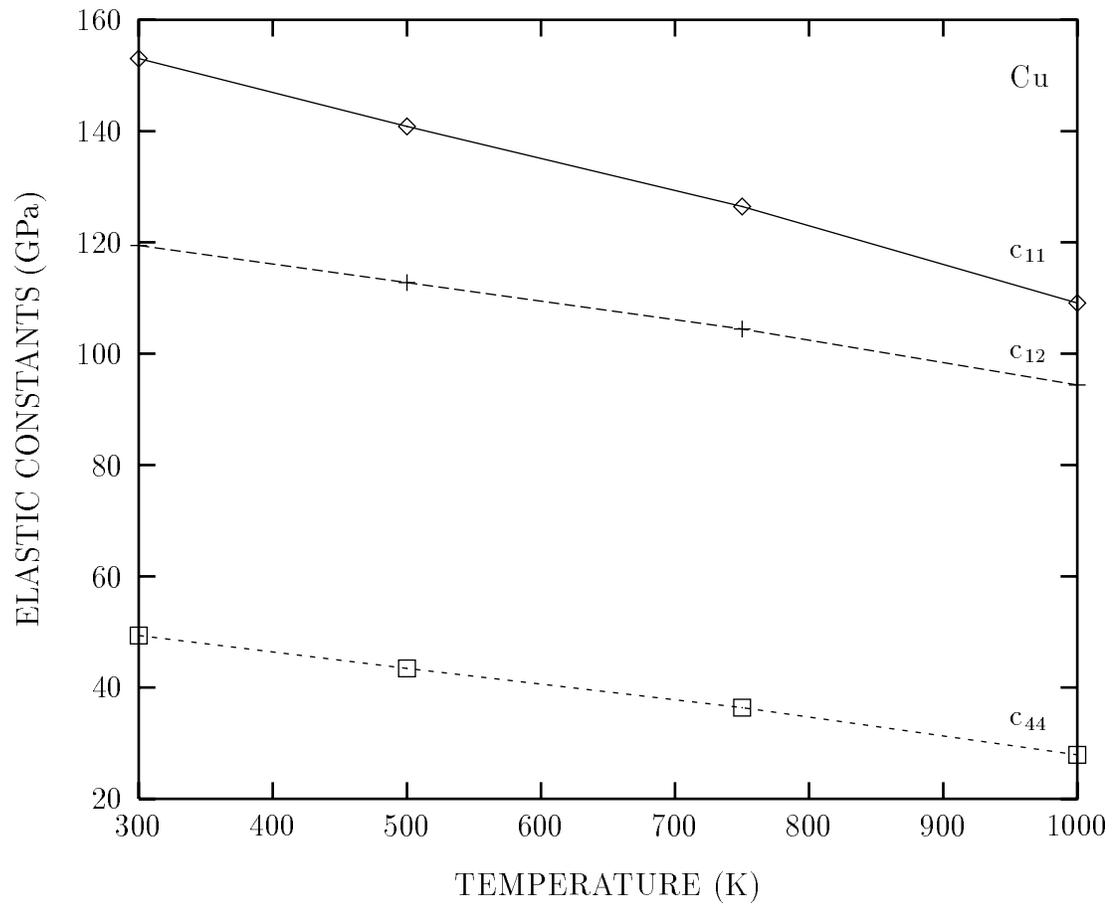
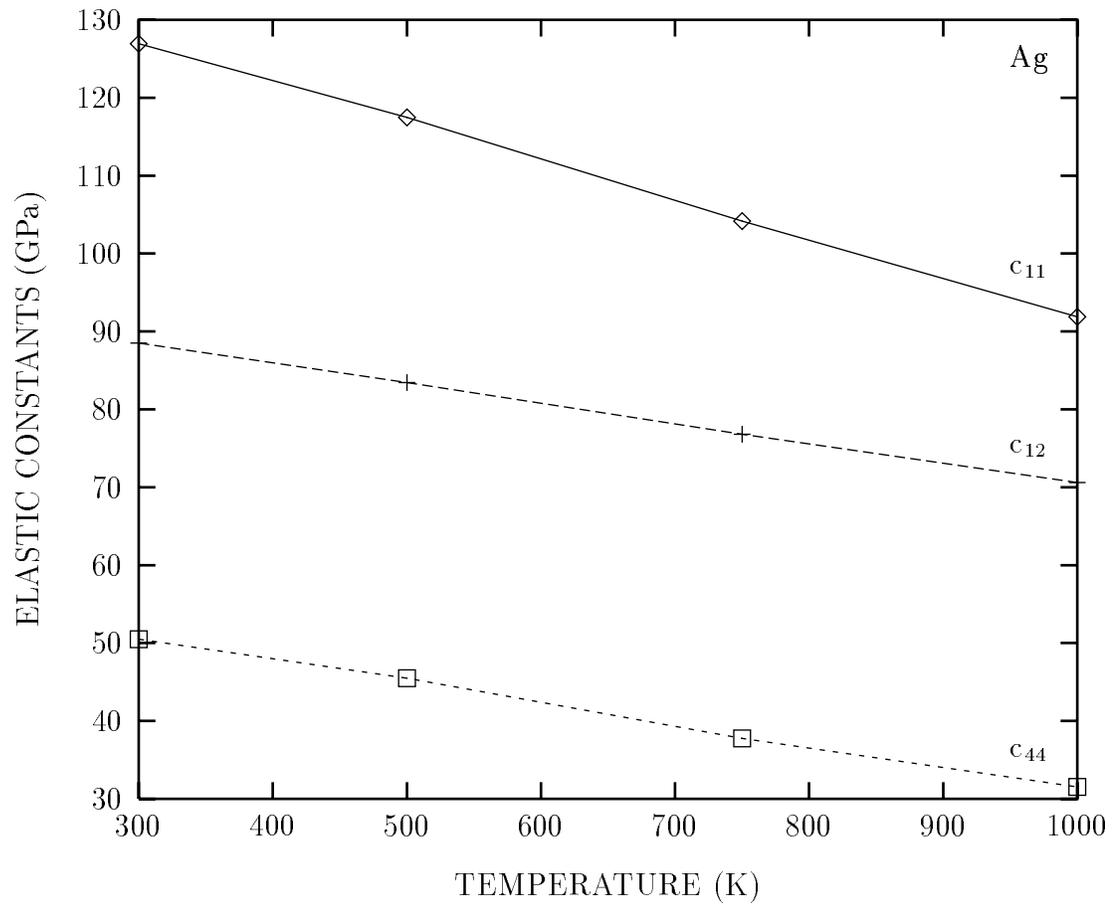
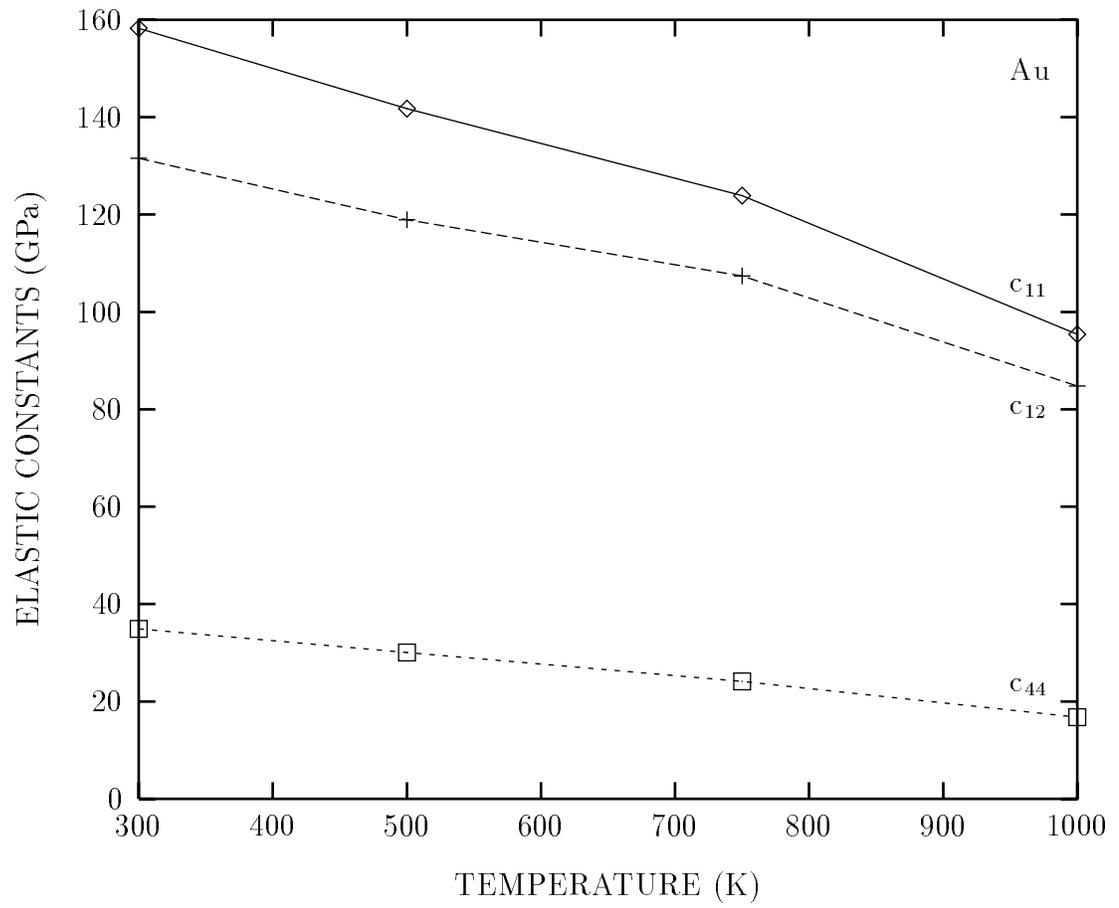
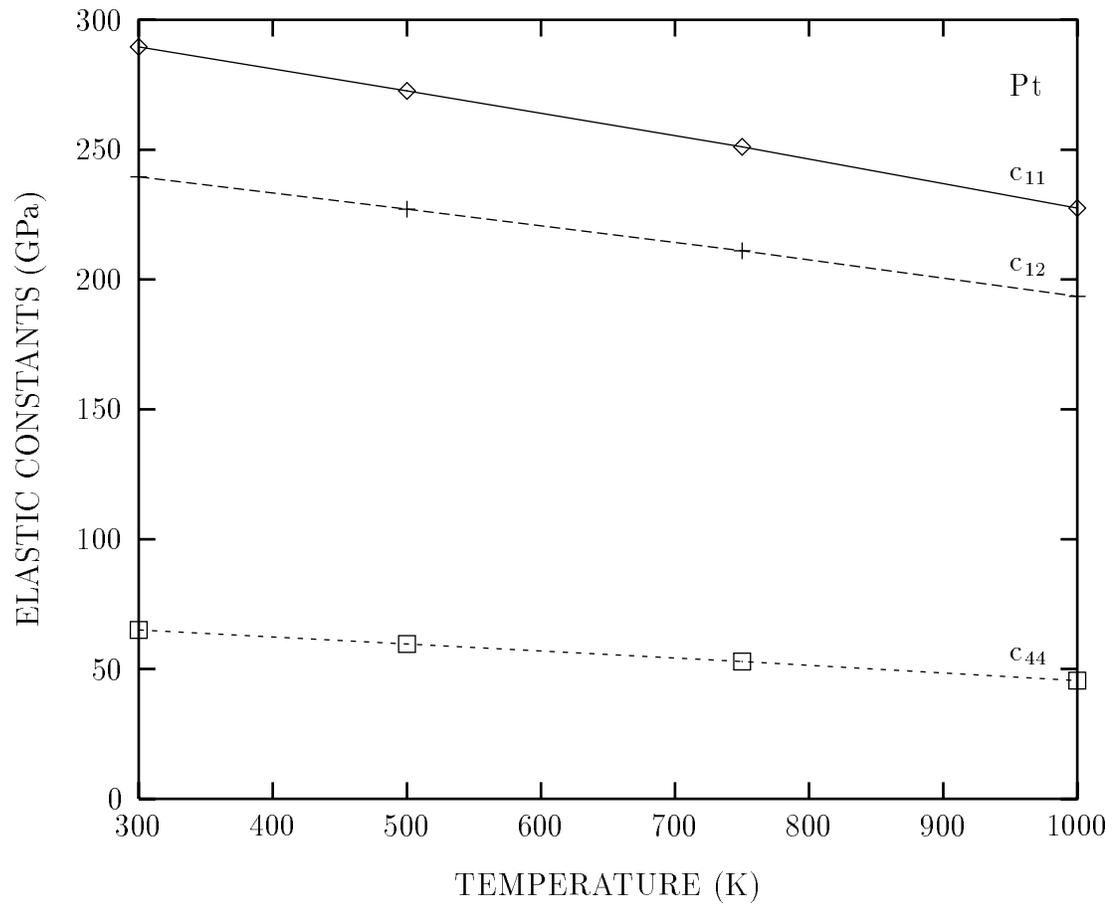
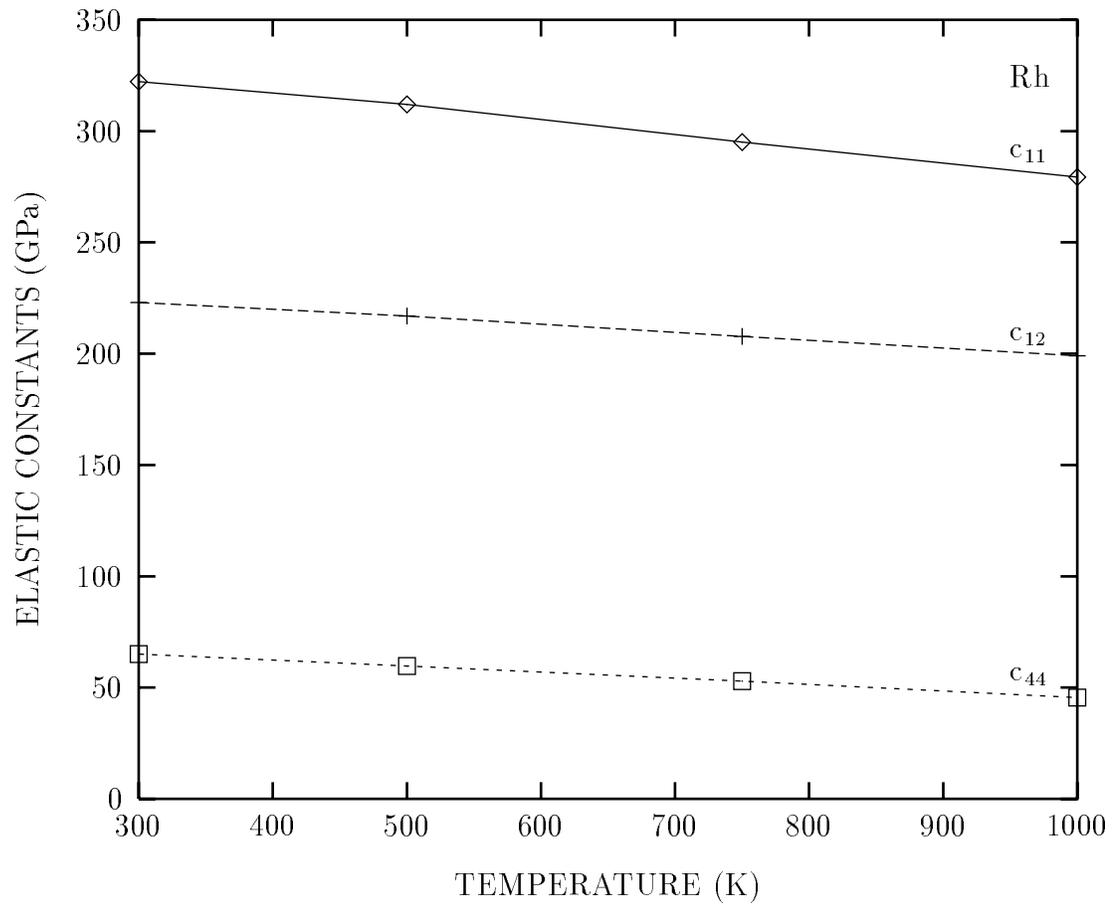
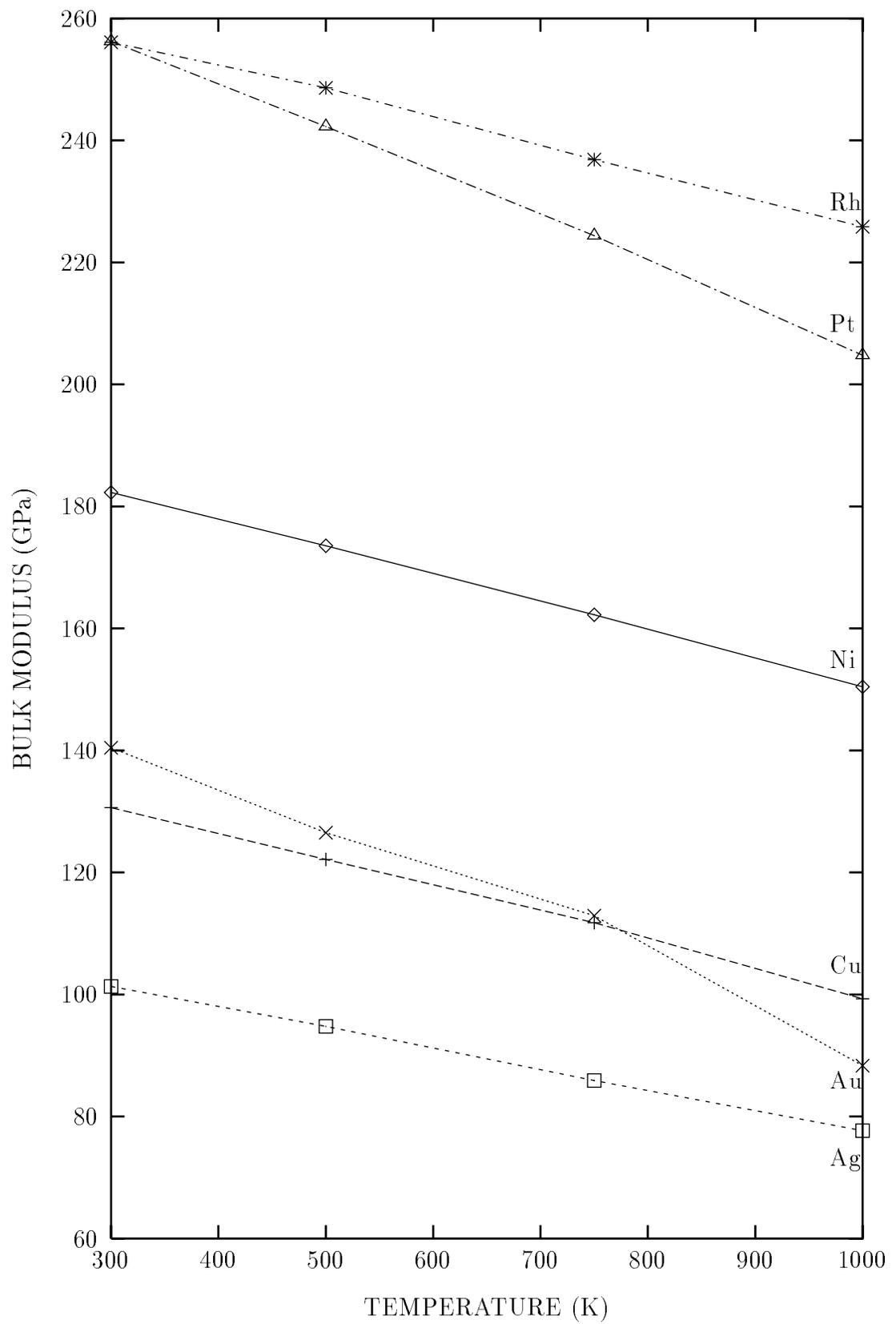
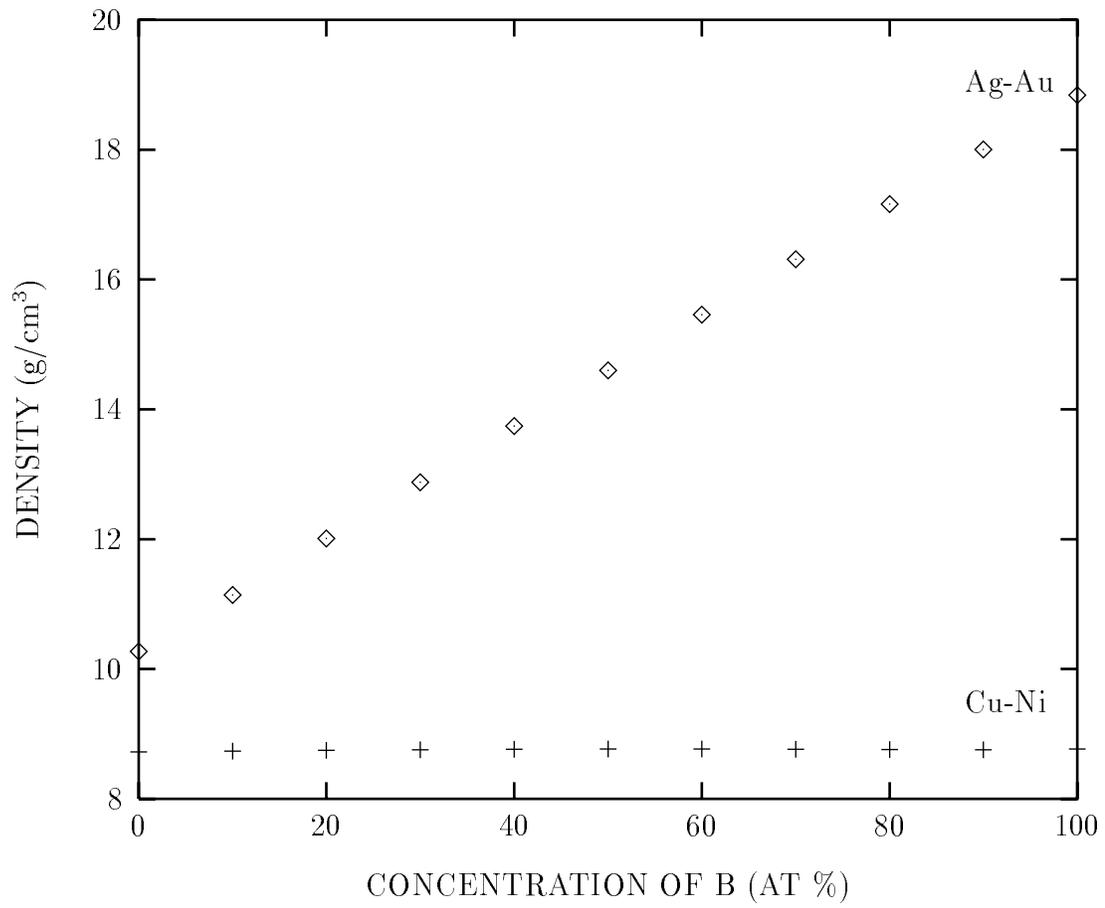
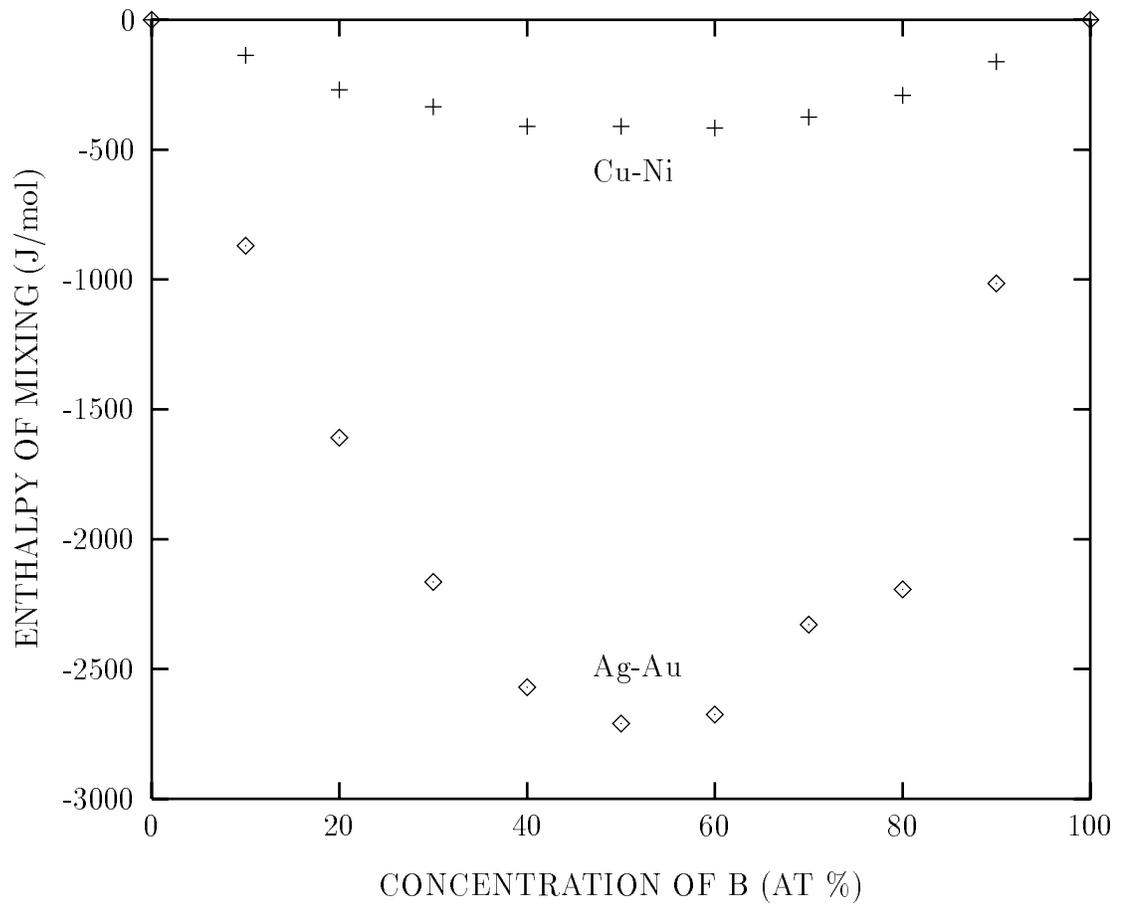
\end{document}